\RequirePackage{lineno}
\documentclass[aps,prd,twocolumn,showpacs,groupedaddress,floatfix,superscriptaddress]{revtex4}  
\usepackage{graphicx}  
\usepackage{dcolumn}   
\usepackage{bm}        
\usepackage{amssymb}   
\usepackage{epsfig}   

\begin{document}

\hspace{5.2in} \mbox{FERMILAB-PUB-12-084-E}

\title{Observation of a narrow mass state decaying into
{\boldmath $\Upsilon(1S) + \gamma$ in $p \overline p$ collisions at  $\sqrt{s}$ = 1.96 TeV}}
%
\affiliation{LAFEX, Centro Brasileiro de Pesquisas F\'{i}sicas, Rio de Janeiro, Brazil}
\affiliation{Universidade do Estado do Rio de Janeiro, Rio de Janeiro, Brazil}
\affiliation{Universidade Federal do ABC, Santo Andr\'e, Brazil}
\affiliation{University of Science and Technology of China, Hefei, People's Republic of China}
\affiliation{Universidad de los Andes, Bogot\'a, Colombia}
\affiliation{Charles University, Faculty of Mathematics and Physics, Center for Particle Physics, Prague, Czech Republic}
\affiliation{Czech Technical University in Prague, Prague, Czech Republic}
\affiliation{Center for Particle Physics, Institute of Physics, Academy of Sciences of the Czech Republic, Prague, Czech Republic}
\affiliation{Universidad San Francisco de Quito, Quito, Ecuador}
\affiliation{LPC, Universit\'e Blaise Pascal, CNRS/IN2P3, Clermont, France}
\affiliation{LPSC, Universit\'e Joseph Fourier Grenoble 1, CNRS/IN2P3, Institut National Polytechnique de Grenoble, Grenoble, France}
\affiliation{CPPM, Aix-Marseille Universit\'e, CNRS/IN2P3, Marseille, France}
\affiliation{LAL, Universit\'e Paris-Sud, CNRS/IN2P3, Orsay, France}
\affiliation{LPNHE, Universit\'es Paris VI and VII, CNRS/IN2P3, Paris, France}
\affiliation{CEA, Irfu, SPP, Saclay, France}
\affiliation{IPHC, Universit\'e de Strasbourg, CNRS/IN2P3, Strasbourg, France}
\affiliation{IPNL, Universit\'e Lyon 1, CNRS/IN2P3, Villeurbanne, France and Universit\'e de Lyon, Lyon, France}
\affiliation{III. Physikalisches Institut A, RWTH Aachen University, Aachen, Germany}
\affiliation{Physikalisches Institut, Universit\"at Freiburg, Freiburg, Germany}
\affiliation{II. Physikalisches Institut, Georg-August-Universit\"at G\"ottingen, G\"ottingen, Germany}
\affiliation{Institut f\"ur Physik, Universit\"at Mainz, Mainz, Germany}
\affiliation{Ludwig-Maximilians-Universit\"at M\"unchen, M\"unchen, Germany}
\affiliation{Fachbereich Physik, Bergische Universit\"at Wuppertal, Wuppertal, Germany}
\affiliation{Panjab University, Chandigarh, India}
\affiliation{Delhi University, Delhi, India}
\affiliation{Tata Institute of Fundamental Research, Mumbai, India}
\affiliation{University College Dublin, Dublin, Ireland}
\affiliation{Korea Detector Laboratory, Korea University, Seoul, Korea}
\affiliation{CINVESTAV, Mexico City, Mexico}
\affiliation{Nikhef, Science Park, Amsterdam, the Netherlands}
\affiliation{Radboud University Nijmegen, Nijmegen, the Netherlands}
\affiliation{Joint Institute for Nuclear Research, Dubna, Russia}
\affiliation{Institute for Theoretical and Experimental Physics, Moscow, Russia}
\affiliation{Moscow State University, Moscow, Russia}
\affiliation{Institute for High Energy Physics, Protvino, Russia}
\affiliation{Petersburg Nuclear Physics Institute, St. Petersburg, Russia}
\affiliation{Instituci\'{o} Catalana de Recerca i Estudis Avan\c{c}ats (ICREA) and Institut de F\'{i}sica d'Altes Energies (IFAE), Barcelona, Spain}
\affiliation{Uppsala University, Uppsala, Sweden}
\affiliation{Lancaster University, Lancaster LA1 4YB, United Kingdom}
\affiliation{Imperial College London, London SW7 2AZ, United Kingdom}
\affiliation{The University of Manchester, Manchester M13 9PL, United Kingdom}
\affiliation{University of Arizona, Tucson, Arizona 85721, USA}
\affiliation{University of California Riverside, Riverside, California 92521, USA}
\affiliation{Florida State University, Tallahassee, Florida 32306, USA}
\affiliation{Fermi National Accelerator Laboratory, Batavia, Illinois 60510, USA}
\affiliation{University of Illinois at Chicago, Chicago, Illinois 60607, USA}
\affiliation{Northern Illinois University, DeKalb, Illinois 60115, USA}
\affiliation{Northwestern University, Evanston, Illinois 60208, USA}
\affiliation{Indiana University, Bloomington, Indiana 47405, USA}
\affiliation{Purdue University Calumet, Hammond, Indiana 46323, USA}
\affiliation{University of Notre Dame, Notre Dame, Indiana 46556, USA}
\affiliation{Iowa State University, Ames, Iowa 50011, USA}
\affiliation{University of Kansas, Lawrence, Kansas 66045, USA}
\affiliation{Kansas State University, Manhattan, Kansas 66506, USA}
\affiliation{Louisiana Tech University, Ruston, Louisiana 71272, USA}
\affiliation{Boston University, Boston, Massachusetts 02215, USA}
\affiliation{Northeastern University, Boston, Massachusetts 02115, USA}
\affiliation{University of Michigan, Ann Arbor, Michigan 48109, USA}
\affiliation{Michigan State University, East Lansing, Michigan 48824, USA}
\affiliation{University of Mississippi, University, Mississippi 38677, USA}
\affiliation{University of Nebraska, Lincoln, Nebraska 68588, USA}
\affiliation{Rutgers University, Piscataway, New Jersey 08855, USA}
\affiliation{Princeton University, Princeton, New Jersey 08544, USA}
\affiliation{State University of New York, Buffalo, New York 14260, USA}
\affiliation{Columbia University, New York, New York 10027, USA}
\affiliation{University of Rochester, Rochester, New York 14627, USA}
\affiliation{State University of New York, Stony Brook, New York 11794, USA}
\affiliation{Brookhaven National Laboratory, Upton, New York 11973, USA}
\affiliation{Langston University, Langston, Oklahoma 73050, USA}
\affiliation{University of Oklahoma, Norman, Oklahoma 73019, USA}
\affiliation{Oklahoma State University, Stillwater, Oklahoma 74078, USA}
\affiliation{Brown University, Providence, Rhode Island 02912, USA}
\affiliation{University of Texas, Arlington, Texas 76019, USA}
\affiliation{Southern Methodist University, Dallas, Texas 75275, USA}
\affiliation{Rice University, Houston, Texas 77005, USA}
\affiliation{University of Virginia, Charlottesville, Virginia 22901, USA}
\affiliation{University of Washington, Seattle, Washington 98195, USA}
\author{V.M.~Abazov} \affiliation{Joint Institute for Nuclear Research, Dubna, Russia}
\author{B.~Abbott} \affiliation{University of Oklahoma, Norman, Oklahoma 73019, USA}
\author{B.S.~Acharya} \affiliation{Tata Institute of Fundamental Research, Mumbai, India}
\author{M.~Adams} \affiliation{University of Illinois at Chicago, Chicago, Illinois 60607, USA}
\author{T.~Adams} \affiliation{Florida State University, Tallahassee, Florida 32306, USA}
\author{G.D.~Alexeev} \affiliation{Joint Institute for Nuclear Research, Dubna, Russia}
\author{G.~Alkhazov} \affiliation{Petersburg Nuclear Physics Institute, St. Petersburg, Russia}
\author{A.~Alton$^{a}$} \affiliation{University of Michigan, Ann Arbor, Michigan 48109, USA}
\author{G.~Alverson} \affiliation{Northeastern University, Boston, Massachusetts 02115, USA}
\author{M.~Aoki} \affiliation{Fermi National Accelerator Laboratory, Batavia, Illinois 60510, USA}
% ...
%
%
\collaboration{The D0 Collaboration\footnote{with visitors from
$^{a}$Augustana College, Sioux Falls, SD, USA,
$^{b}$The University of Liverpool, Liverpool, UK,
$^{c}$UPIITA-IPN, Mexico City, Mexico,
$^{d}$DESY, Hamburg, Germany,
,
$^{e}$SLAC, Menlo Park, CA, USA,
$^{f}$University College London, London, UK,
$^{g}$Centro de Investigacion en Computacion - IPN, Mexico City, Mexico,
$^{h}$ECFM, Universidad Autonoma de Sinaloa, Culiac\'an, Mexico
and
$^{i}$Universidade Estadual Paulista, S\~ao Paulo, Brazil.
$^{\ddag}$Deceased.
}} \noaffiliation
\vskip 0.25cm
\date{March 26, 2012}

\begin{abstract}
Using data corresponding to an integrated luminosity of 1.3 fb$^{-1}$, we observe a narrow mass state decaying into $\Upsilon(1S)+\gamma$, where the $\Upsilon(1S)$ meson is  detected 
by its decay into a pair of oppositely charged muons, and the photon is identified through its conversion into an electron-positron pair. The significance of this
observation is 5.6 standard deviations. The mass of the state is centered at $10.551 \pm 0.014 \mathrm{(stat.)}\pm 0.017\mathrm{(syst.)}$~GeV/$c^2$, which is consistent with that of the state recently observed by the ATLAS Collaboration. 
\end{abstract}

\maketitle 

The study of heavy quarkonium is of fundamental importance to our understanding of perturbative and non-perturbative
quantum chromodynamics (QCD). The heavy quarkonium state is one of the simplest QCD systems and the
observed masses and branching fractions are in quantitative agreement with
theoretical expectations for almost all of the known states. The recent
observations of unexpected quarkonium-like states have, however, raised new questions.
The discovery of the $X(3872)$ in 2003~\cite{Choi:2003ue} provided the first surprise.  
One explanation for this
state is a $DD^*$ molecule~\cite{Tornqvist:2004qy}, but the subsequent discoveries
of several other unexpected particles containing a $c\overline{c}$ pair have suggested otherwise.
The $Y(4260)$~\cite{Aubert:2005rm} is most naturally explained as a hybrid $c\overline{c}$-gluon 
state~\cite{Zhu:2005hp},
but it could also be a four-quark state~\cite{Maiani:2005pe}. 
 The $Z(4430)$~\cite{:2007wga} is likely a four-quark charmonium-like particle by virtue of its non-zero electric charge, and the recently observed charged structures decaying into $\pi\Upsilon$ might be similar states containing $b$ quarks~\cite{Belle:Zb}.   

Discoveries of more $b$-quark counterparts to these exotic states could
shed light on the underlying structure of this class of particles. In this Letter, we present a search for such new particles
decaying into $\Upsilon(1S)+\gamma$. This decay mode is shared by the $P$-wave $\chi_b(1P)$ and $\chi_b(2P)$  bottomonium states, which each consist of three spin states ($J = 0, 1, 2$) with hyperfine mass splittings.  In addition to these known particles, the ATLAS Collaboration has recently published the observation of a narrow structure in this decay mode with a mass of $10.530 \pm 0.005 \mathrm{(stat.)} \pm 0.009 \mathrm{(syst.)}$~GeV/$c^2$, which they interpret as the $\chi_b(3P)$ system~\cite{Aad:2011ih}.

\begin{figure}
\centerline{\epsfig{figure=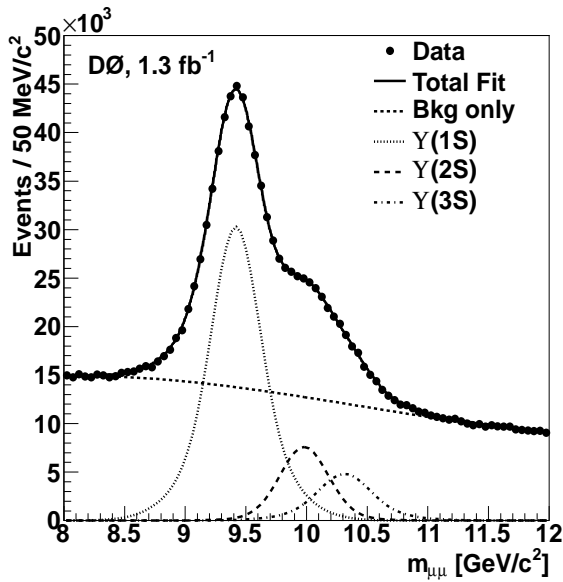,width=0.45\textwidth,height=0.45\textwidth}}
\caption{Dimuon invariant mass spectrum for opposite-charge pairs passing the muon selection criteria. The solid curve is a fit to the data assuming three $\Upsilon$ resonances and a combinatorial background. The relative contributions from the $\Upsilon(1S)$, $\Upsilon(2S)$, and $\Upsilon(3S)$ states are also shown.}
\label{fig:dimuon}
\end{figure}

\begin{figure}
\centerline{\epsfig{figure=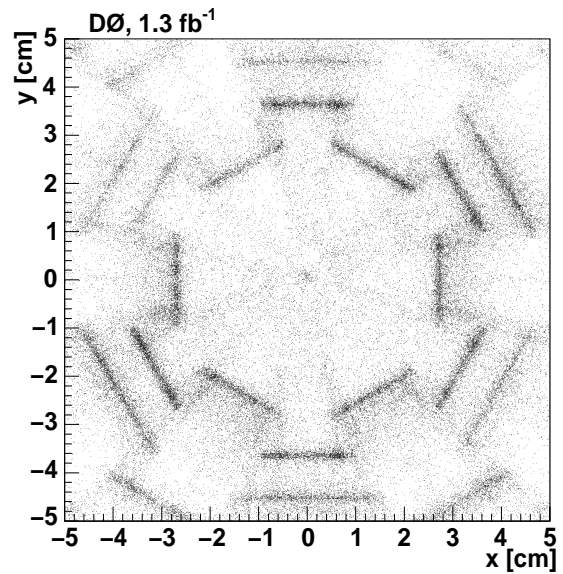,width=0.45\textwidth,height=0.45\textwidth}}
\caption{The vertex position in the $x-y$ plane for photon conversion candidates passing 
the photon selection requirements.  The $x-y$ plane 
is perpendicular to the beam, with $y$ pointing upwards and positive $x$ pointing to the right when viewed in the
anti-proton direction.}
\label{fig:xray}
\end{figure}

\begin{figure*}[t!]
\epsfig{figure=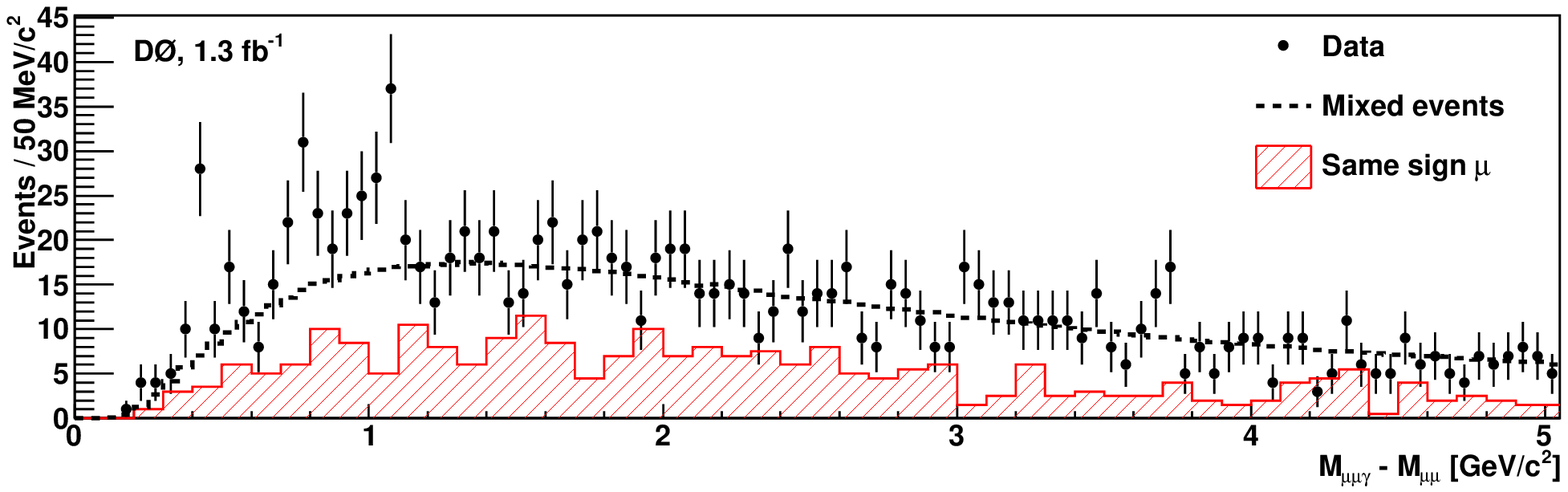,width=\textwidth}

\caption{Mass difference $M_{\mu\mu\gamma} - M_{\mu\mu}$ for events passing all selection criteria. 
The curve shows the mass difference for the background model which combines 
$\Upsilon(1S)$ and $\gamma$ candidates from different events, normalized to the number of
data events above 1.2 GeV/$c^2$. The hatched area shows the distribution obtained by repeating the event selection using muons with the same charge instead of those of opposite charge.}
\label{fig:peakandmixed}
\end{figure*}

The data used for this analysis corresponds to 1.3~fb$^{-1}$ of 
integrated luminosity collected with the D0 experiment between April 2002 and February 2006.
The D0 detector is described in detail elsewhere~\cite{d0det}. The 
detector elements crucial for this search are the central tracking and muon systems. The 
central tracking system consists of a silicon microstrip tracker (SMT) and a central fiber tracker (CFT),
both located within a 2~T superconducting solenoidal
magnet with pseudorapidity coverage $|\eta|<3$ and $|\eta|<2.5$, respectively.
The muon system, which extends to  $|\eta| \approx 2$, consists of a layer of tracking 
detectors and scintillation trigger  counters in front of 1.8~T iron toroids, followed 
by two similar layers after the toroids. The events used in this analysis were collected with 
an inclusive mixture of single muon and dimuon triggers. 

The event selection for this analysis is chosen to maximize the signal significances of the known $\chi_b (1P)$  and $\chi_b (2P)$ states, as well as those of the charmonium $\chi_c$ states which are observed in the analogous $J/\psi + \gamma$ decay mode. The search region above the $\chi_b(2P)$ mass and below the $B\overline{B}$ mass threshold (10.38 - 10.63 GeV/$c^2$) is not examined in this optimization procedure. 
Muon candidates are required to be reconstructed in both the inner and outer layers of the muon detector and must be matched to tracks in
the central tracking system with $p_T  > 1.5$ GeV/$c$. The tracks are required to originate from a common $p \overline{p}$ interaction vertex. The resulting mass distribution 
for muon pairs of opposite charge is shown in Fig.\ \ref{fig:dimuon}. We select events in the 
mass range 9.1 GeV/$c^2 < M_{\mu\mu} < 9.7$ GeV/$c^2$, which contains approximately 275,000 
$\Upsilon(1S)$ candidates after background subtraction. 

Photons are detected via their conversions into electron-positron pairs. 
We use this technique instead of direct detection in the calorimeter because the energies of photons from quarkonia decay are typically too low to be precisely measured in the calorimeter, while the tracking system provides excellent resolution in this kinematic region. This allows for a separation of closely-spaced mass states.  Well-measured track pairs of opposite charge that form a good vertex are selected. These tracks are required to be loosely associated to the same $p \overline{p}$ interaction vertex as the muon tracks, but since photons travel some distance before they are converted in the material of the detector, the electron-positron tracks are required to originate away from that vertex. 
The two tracks are combined to form a photon candidate. The resulting photon trajectory is required to intersect the beam axis within 0.8 cm of where each of the  tracks intersect it. The momentum vector of the photon must also point back from its
reconstructed conversion point to the $p \overline{p}$ collision vertex, and the invariant mass of the conversion pair must be $M_{ee} < 80$ MeV$/c^2$.  These selection criteria are confirmed by a study of double conversion pairs from $\pi^0 \rightarrow \gamma \gamma$ decays. The origin, in the plane transverse to the beam direction, of the track pairs passing these criteria is 
shown in Fig.\ \ref{fig:xray}, where the structure of the silicon tracker is clearly visible. 

Converted photons passing these requirements are combined with muons passing the $\Upsilon(1S)$ selection. The photon and both muons must be consistent with coming from a common vertex, 
and both muon trajectories must intersect the beam axis within a distance of 1.2\ cm from that of the photon.  
The $\Upsilon(1S) \gamma$ system is also required to have $p_T >$ 5 GeV/$c$. 

The resulting distribution of $\Delta M = M_{\mu\mu\gamma} - M_{\mu\mu}$ is shown in Fig.\ \ref{fig:peakandmixed}. 
Peaks corresponding to the $\chi_b(1P)$ and $\chi_b(2P)$ states are clearly seen at $\Delta M \approx 0.4$ GeV/$c^2$ and $\Delta M \approx 0.8$ GeV/$c^2$, respectively.
A third peak is also observed centered around $\Delta M \approx$ 1 GeV/$c^2$, which is consistent with the recent observation of a new state by the ATLAS  Collaboration. The mass resolution is not good enough to separate the hyperfine splitting of the known $\chi_b$ states, and we find no indication of sub-structure in the mass region of the new state. The measured values $\Delta M_ {\chi_b(1P)} = 0.418 \pm 0.005$ GeV/$c^2$ and $\Delta M_ {\chi_b(2P)} = 0.760 \pm 0.014$ GeV/$c^2$ are shifted from their true values due to energy loss of the electron/positrons. A scale factor of  $0.96 \pm  0.01$ is determined by comparing these measurements to their world average values assuming an equal mixture of $J=1$ and $J=2$ components for each $\chi_b$ state (the $J=0$ components are suppressed in this decay mode)~\cite{PDG}. The measured masses of the $\chi_c$ states and the $\pi^{0}$ detected using photon conversions have a shift consistent with this scale factor.

The shape of the background distribution is determined from the data by combining $\Upsilon(1S)$ and photon candidates from different events. As seen in Fig.\ \ref{fig:peakandmixed}, this mixed event background model describes the data for 
a wide range of $\Delta M$ outside the region of interest. We also study the  $\Delta M$  distributions for events with dimuons in the $\Upsilon(1S)$ mass sideband regions and for events with dimuons with the same charge in the $\Upsilon(1S)$ mass region. The resulting $\Delta M$ distributions for these selections show no peaking structure and have shapes similar to that of the mixed-event background model. 

The mass distribution $M = M_{\mu\mu\gamma} - M_{\mu\mu} + m_{\Upsilon(1S)}$, where $m_{\Upsilon(1S)}$ is the world average value 9.4603 GeV/$c^2$~\cite{PDG}, is shown in Fig.~4 along with the results of  an unbinned maximum likelihood fit with three signal peaks and a background shape determined from the mixed event model.  Crystal Ball functions~\cite{CBall} are used to describe the signal mass shapes to take into account the radiative tails due to bremsstrahlung. We use single Crystal Ball functions to describe the  mass of the $\chi_b(1P)$ and $\chi_b(2P)$ systems and that of the new state. The center for each $\chi_b$ mass function is fixed to its world average value corrected by the electron/positron energy loss scale factor. The widths of the signal functions are described by a single parameter scaled by the mass of each state, and  the lengths of the radiative tails are the same for all three states. These constraints, determined from the data without consideration of the new structure, have also been verified using Monte Carlo simulations.  The widths of all three peaks obtained in the fit are compatible with the D0 detector's resolution. The fit yields $65 \pm 11$ events above background corresponding to the new state. A similarly good fit is also obtained by using an exponential function multiplied by a low-mass turn on curve to describe the background. The shape of the resulting background agrees well with that of the mixed-event model.

\begin{figure}[t]
\epsfig{figure=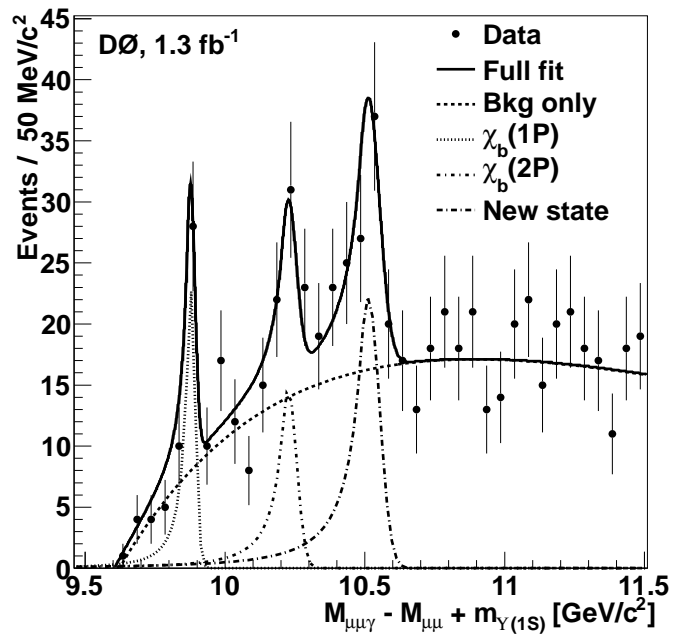,width=0.5\textwidth}
\caption{The distribution of $M = M_{\mu\mu\gamma} - M_{\mu\mu} + m_{\Upsilon(1S)}$ fit with three signal functions and the mixed event background.}
\label{fig:peakfitted}
\end{figure}

A significance of more than six standard deviations is determined from the difference in the log likelihood of the fits with and without the new state's contribution. Considering the probability of an upward fluctuation of the background producing a signal of this width anywhere in the search
region, reduces the significance to 5.6 standard deviations. The mass of the new state is corrected by the same scale factor used to fit the $\chi_b(1P, 2P)$ states and is measured to be  $10.551 \pm 0.014 \mathrm{(stat.)} \pm 0.017 \mathrm{(syst.)}$~GeV/$c^2$. The main sources of systematic uncertainty are due to the unknown mixture of $\chi_b$ spin states ($13$ MeV/$c^2$), the mass scale correction ($10$ MeV/$c^2$), and variations in the background model ($5$ MeV/$c^2$). 

In summary, we present a search for new particles with masses below the $B\overline{B}$ threshold which decay into  $\Upsilon(1S) + \gamma$. In addition to the known states $\chi_b(1P)$ and $\chi_b(2P)$, a third 
peak is observed at a mass consistent with the new state observed by the ATLAS collaboration. No other mass peaks are observed. The background distribution is well described over a wide mass range by a background model which mixes $\Upsilon(1S)$ and $\gamma$ candidates from different events. A likelihood fit to the mass distribution results in a measured mass of $10.551 \pm 0.014 \mathrm{(stat.)}\pm 0.017\mathrm{(syst.)}$ GeV/$c^2$ for the new state with a width consistent with the D0 detector's mass resolution.  Further analysis is underway to determine whether this structure is due to the  $\chi_b(3P)$ system or some exotic bottom-quark state.

%
We thank the staffs at Fermilab and collaborating institutions,
and acknowledge support from the
DOE and NSF (USA);
CEA and CNRS/IN2P3 (France);
MON, Rosatom and RFBR (Russia);
CNPq, FAPERJ, FAPESP and FUNDUNESP (Brazil);
DAE and DST (India);
Colciencias (Colombia);
CONACyT (Mexico);
NRF (Korea);
FOM (The Netherlands);
STFC and the Royal Society (United Kingdom);
MSMT and GACR (Czech Republic);
BMBF and DFG (Germany);
SFI (Ireland);
The Swedish Research Council (Sweden);
and
CAS and CNSF (China).
%

\end{document}